\documentclass[mathleft,cite,graphics,graphicx,psfig,amssymb]{an}
\usepackage{graphicx}
\usepackage{amsfonts}
\usepackage{natbib}
\usepackage{times}
\begin{document}
\Pagespan{1}{}%
\Yearpublication{2008}%
\Yearsubmission{2008}%
\Month{9}%
\Volume{999}%
\Issue{88}%
\DOI{DOI}%

\title{New absolute magnitude calibrations for detached binaries}

\author{S. Bilir\inst{1}\fnmsep\thanks{Corresponding author: \email{sbilir@istanbul.edu.tr}\newline} 
\and T. Ak\inst{1,2}
\and E. Soydugan\inst{3}
\and F. Soydugan\inst{3}
\and E. Yaz\inst{1}
\and N. Filiz Ak \inst{4}
\and Z. Eker\inst{2,3}
\and O. Demircan\inst{3}
\and M. Helvac\i \inst{5}
}

\institute{Istanbul University, Faculty of Sciences, Department of Astronomy and 
Space Sciences, 34119 Istanbul, Turkey
\and
T\"UB\.ITAK National Observatory, Akdeniz University Campus, 07058 Antalya, 
Turkey
\and
\c Canakkale Onsekiz Mart University, Faculty of Sciences and Arts, Department 
of Physics, 17100 \c Canakkale, Turkey 
\and
Erciyes University, Faculty of Sciences and Arts, Department of Astronomy and 
Space Sciences, Talas Yolu, 38039 Kayseri, Turkey
\and
Ankara University, Faculty of Sciences, Department of Astronomy and 
Space Sciences, 06100 Ankara, Turkey
}
\date{} 

\received{}
\accepted{}
\publonline{later} 
\keywords{stars: distances, (stars:) binaries: eclipsing}

\abstract{Lutz-Kelker bias corrected absolute magnitude calibrations for the detached 
binary systems with main-sequence components are presented. The absolute magnitudes of the 
calibrator stars were derived at intrinsic colours of Johnson-Cousins and {\em 2MASS} 
(Two Micron All Sky Survey) photometric systems. As for the calibrator stars, 
44 detached binaries were selected from the {\em Hipparcos} catalogue, which have 
relative observed parallax errors smaller than 15\% ($\sigma_{\pi}/\pi\leq0.15$). 
The calibration equations which provide the corrected absolute 
magnitude for optical and near-infrared pass bands are valid for wide ranges of colours 
and absolute magnitudes: $-0.18<(B-V)_{0}<0.91$, $-1.6<M_{V}<5.5$ and 
$-0.15<(J-H)_{0}<0.50$, $-0.02<(H-K_{s})_{0}<0.13$, $0<M_{J}<4$, respectively. 
The distances computed using the luminosity-colours (LCs) relation with optical ($BV$) and 
near-infrared ($JHK_{s}$) observations were compared to the distances found from 
various other methods. The results show that new absolute magnitude calibrations of 
this study can be used as a convenient statistical tool to estimate the true 
distances of detached binaries out of {\em Hipparcos'} distance limit.}

\maketitle

\section{Introduction}
The most reliable physical parameters of stars, e.g. masses, radii, effective 
temperatures etc., used in testing theoretical predictions of fundamental 
physical structure of single stars, are obtained from the observations of 
detached eclipsing binaries with main-sequence components. Masses and radii 
could be determined within an accuracy better than 1\% from their light and 
radial velocity curves \citep{Andersen1991, Southworthetal2004, Southworthetal2005a}.

There could be several techniques to determine distances for detached eclipsing 
binaries. The most reliable distances are usually the ones from 
accurately measured trigonometric parallaxes or interferometric 
observations of nearby stars. Relatively less reliable distances are the ones 
based on spectroscopic or photometric parallaxes using absolute magnitudes 
estimated or computed from the solutions of data obtained by photometric or 
spectroscopic observations. Such distances, however, require measurements of reliable 
reddening free apparent magnitudes. The effect of interstellar reddening on 
the final distance predictions could be larger than expected. Since reddening 
is less effective at infrared wavelengths, the prediction errors could be 
minimized by using infrared photometry \citep{Southworthetal2005b}.

Relation between orbital period, luminosity and colours of the binary 
stars, called PLC relation, was first proposed by \cite{Rucinski1974} 
\citep[see also,][]{Mochnacki1981} for W UMa-type binaries in order to reduce 
the scatter in the main-sequence colour-magnitude relation. \cite{Rucinski1994} 
established the PLC relation of W UMa-type stars using ($B-V$), ($V-I_{c}$) 
colours and orbital periods. The PLC relation of W UMa-type binaries was 
improved later using {\em Hipparcos} parallaxes \citep{RucinskiandDuerbeck1997}. 
In a recent paper, \cite{Aketal2007} described a similar relation, called 
PLCs relation for cataclysmic binaries. Although it is possible to find 
a PLCs relation for detached binaries too, applying this idea to detached 
binaries makes no sense since the orbital period is not correlated in any 
physical way to the components. Thus, a relation between the luminosity 
(absolute magnitude) and colours can be more useful for detached binaries. 
It should be noted that more accurate calibrations can be obtained by using 
two colours instead of one colour, since one colour dependent calibrations 
may include systematic errors \citep[see, e.g.][]{Biliretal2008}.

In this study, we first estimate the systemic $V$ and $J$ band absolute magnitudes 
$M_{V}$ and $M_{J}$ of detached binaries using reliable trigonometric parallaxes 
from {\em Hipparcos} satellite by assuming that the light in $V$ and $J$ bands 
comes from the system as a $\it whole$. Then, we find the dependence of the
absolute magnitude on de-reddened colours $(B-V)_{0}$, $(J-H)_{0}$ and 
$(H-K_{s})_{0}$ to derive absolute magnitude calibrations for detached binaries 
with observations in Johnson-Cousins and {\em 2MASS} photometric systems. 
We should state that our aim is $\it not$ to measure the absolute magnitudes 
of each component in the detached binaries in our list. 

Unfortunately, measured parallaxes are never free from systematic errors. This 
problem has been noticed by \citet[LK,][]{LK73}. Assuming uniform spatial 
distribution of stars and a uniform distribution of observed parallaxes around 
the true parallax, LK explained that there must be a systematic error, known as 
Lutz-Kelker bias, in computed distances which depends only upon relative parallax 
error ($\sigma_{\pi}/\pi$), where $\pi$ is observed parallax and $\sigma_{\pi}$ is 
the associated standard error. Because of LK bias, the true parallaxes are always 
smaller than observed parallaxes \citep{LK73}. Therefore observed parallaxes are 
needed to be corrected. This correction procedure is recognized as LK correction. 

According to \cite{J01}, only the studies which are careful to use parallaxes with 
($\sigma_{\pi}/\pi)\leq0.1$ could be excused since the bias could be tolerable. Otherwise, 
LK bias (if not corrected) would either alter conclusions or invalidate the results. 
However, the standard LK corrections become significant if ($\sigma_{\pi}/\pi)\geq0.05$ 
and increase as relative error gets bigger \citep{M05}. Thus the parallaxes with such 
significant errors must be corrected.

Aim of this study is to establish an absolute magnitude calibration for practical 
usage to estimate distances of detached main-sequence eclipsing binaries which have 
no tri-gonometric parallax nor known physical structures to imply their absolute 
brightness, but only have at least an eclipsing light curve indicating they are on 
the main sequence and detached. Unlike the common practice of predicting binary 
distances photometrically, which requires estimated absolute magnitudes of each 
component extracted from the light contributions of each component and/or their 
physical parameters from a spectroscopic or a light curve solution, this new method 
does not require extensive work; thus practical and quick, therefore, useful especially 
when studying large numbers of eclipsing data such as ASAS \citep{Paczynski06} 
survey where light curve solutions are not available to most binaries.       

\section{The Data}
Photometric data of 44 detached eclipsing binaries with main-sequence components, 
which have observed trigonometric parallaxes ($\pi$) with relative errors smaller than 
($\sigma_{\pi}/\pi)\leq0.15$, were collected from the literature. The 14 of 
them are chromospherically active binaries taken from \cite{Ekeretal2008}. 
Since there are relative errors of the observed parallaxes bigger than 0.05, LK correction 
needs to be applied to derive the true parallaxes ($\pi_{0}$) using following 
equation \citep{Smith1987}:

\begin{equation}
\pi_{0}=\pi(\frac{1}{2}+\frac{1}{2}\sqrt{1-16(\sigma_{\pi}/\pi)^{2}}~).
\end{equation}

Table 1 contains the basic data required for absolute magnitude calibration. 
The columns are organized as order, name, spectral type, galactic coordinates 
($\ell$, $b$), observed parallaxes ($\pi$), relative parallax error ($\sigma_{\pi}/\pi$), 
LK corrected true parallaxes ($\pi_{0}$), colour excess ($E(B-V)$), visual 
brightness ($V$) and colour ($B-V$), infrared brightness ($J$) and colours 
($J-H$, $H-K_{s}$) and quality flag. Quality flag ``AAA'' indicates the best quality 
or maximum reliability of {\em 2MASS} photometric data.

Parallaxes, parallax errors, galactic coordinates and $J$, $H$ and $K_{s}$ 
magnitudes are taken from the {\em Hipparcos} catalogue \citep{ESA97} and the 
Point-Source Catalogue and Atlas \citep{Cutrietal2003} which is based on 
the {\em 2MASS} (Two Micron All Sky Survey) observations, respectively, by using 
VizieR\footnote{http://vizier.u-strasbg.fr/viz-bin/VizieR} service. The 
{\em 2MASS} photometric system comprises Johnson's $J$ (1.25 $\mu$m) and $H$ 
(1.65 $\mu$m) bands with the addition of $K_{s}$ (2.17 $\mu$m) band, which is 
bluer than Johnson's $K$-band \citep{Skrutskieetal2006}. Spectral types were 
collected from the available literature.

\subsection{Intrinsic colours and absolute magnitudes}
In principle, a calibration in any photometric system has to be based on the 
intrinsic (de-reddened) colours, such as $(B-V)_{o}$, $(J-H)_{o}$ and 
$(H-K_{s})_{o}$. However, unlike the observed colours, the intrinsic colours 
require one more step to be estimated from any kinds of colour excess values if 
spectroscopic observations are not available to reveal them independently. For 
estimating the intrinsic colours, the colour excess of $E(B-V)$ has been 
preferred for this study. Unfortunately, only limited number of $E(B-V)$ 
colour excesses of some systems were given in the literature. Therefore, we 
used \cite{Schlegeletal1998} maps and NASA Extragalactic 
Database\footnote{http://nedwww.ipac.caltech.edu/forms/calculator.html} to calculate 
the $E(B-V)$ value of a star. \cite{Schlegeletal1998} maps provide only the 
$E(B-V)$ values according to galactic coordinates, which are the modeled values 
for any direction from Sun to the edges of our Galaxy as a consequence of the 
galactic dust. The colour excess values according to galactic latitude ($b$) 
and longitude ($\ell$) towards the directions of stars are shown by 
$E_{\infty}(B-V)$, which symbolically means up to infinity but actually 
up to the edge of the Galaxy. Therefore, the $E_{\infty}(B-V)$ values have to 
be reduced according to the actual distance of each star. The total interstellar 
absorption within the Galaxy in the photometric $V$ band was computed from an 
available modeled value of $E_{\infty}(B-V)$ for a given galactic latitude ($b$) 
as

\begin{equation}
A_{\infty}(b)=3.1E_{\infty}(B-V).
\end{equation}
The total interstellar absorption in the $V$ band up to the distance ($d=1/\pi$) of 
the star can be estimated by \citep{BahcallandSoneira1980}

\begin{equation}
A_{d}(b)=A_{\infty}(b)\Biggl[1-exp\Biggl(\frac{-\mid d~sin(b)\mid}{H}\Biggr)\Biggr],
\end{equation}
where $H$ is the scaleheight for the interstellar dust which is adopted as 125 pc 
\citep{Marshall2006}. Finally, the colour excess for a star at the distance $d$ is 
estimated from

\begin{equation}
E_{d}(B-V)=A_{d}(b)~/~3.1.
\end{equation}

Once we obtained the colour excess for a star, we have calculated the intrinsic 
magnitudes and colours as following: $V_{o}=V-3.1\times E_{d}(B-V)$, 
$(B-V)_{o}=(B-V)-E_{d}(B-V)$, $J_{o}=J-0.887\times E_{d}(B-V)$, 
$(J-H)_{o}=(J-H)-0.322\times E_{d}(B-V)$, $(H-K_{s})_{o}=(H-K_{s})-0.183\times E_{d}(B-V)$. 
We used the equations of \cite{FiorucciandMunari2003} \citep[see also,][]{BilirGuverAslan2006} 
for the determination of the total absorption for $J$ band and colour excesses 
of ($J-H$) and ($H-K_{s}$) colours. All the magnitudes and colours with 
subscript ``0'' will be mentioned as de-reddened ones, hereafter.

In addition to intrinsic magnitudes and colours, fundamental input data required for 
the calibration are absolute magnitudes. The absolute magnitude of stars in Table 1 
have been computed by the well known distance-modulus formula both for optical and 
near-infrared bands, i.e. $M_{V_{LK}}=V_{o}-5\log (1/\pi_{0})+5$ and 
$M_{J_{LK}}=J_{o}-5\log (1/\pi_{0})+5$, respectively. The computed values of $M_{V}$ 
and $M_{J}$ are given in columns 3 and 5 in Table 2. Corresponding propagated errors 
were calculated as $\delta M=2.17(\sigma_{\pi}/\pi)+ \delta m$, where ($\sigma_{\pi}/\pi$) 
and $\delta m$ are relative parallax error and the error of the apparent magnitude in 
the relevant photometric system, respectively. 

\begin{table*}
\setlength{\tabcolsep}{2pt}
\caption{The data used in calibration of the absolute magnitudes. $\pi_{0}$ shows LK corrected parallax. 
``Quality'' denotes the quality flag of {\em 2MASS} data.}
{\scriptsize
\begin{tabular}{cllcccccccccccc}
\hline
ID &  Star & Spectral & $\ell$ & $b$ & $\pi$ & $(\sigma_{\pi}/\pi)$ & $\pi_{0}$ & $E_{d}(B-V)$ & $V$ & $(B-V)$ & $J$ &   $(J-H)$ & $(H-K_{s})$ &  Quality \\
   &       & type  & ($^{\circ}$) & ($^{\circ}$) & (mas) &     & (mas) & (mag) & (mag) & (mag) &  (mag) & (mag) & (mag) & \\
\hline
 1 &     RT And &    G0V + K2V &    108.058 &     -6.926 &   13.26 &       0.09 &   12.87 &  0.020$^{a}$ & 8.97$\pm$0.017 & 0.546$\pm$0.018 & 8.037$\pm$0.018 & 0.257$\pm$0.027 & 0.093$\pm$0.028 &        AAA \\
 2 &   V805 Aql &    A2 + A7 &     24.163 &     -8.544 &    5.80 &       0.15 &     5.22&  0.130$^{b}$ & 7.60$\pm$0.010 & 0.286$\pm$0.013 & 6.970$\pm$0.027 & 0.079$\pm$0.050 & 0.078$\pm$0.047 &        AAA \\
 3 &  $\beta$ Aur &  A1V + A1V &    167.457 &     10.409 &    39.72 &       0.02 &    39.66& 0.000$^{c}$ & 1.90$\pm$0.005 & 0.077$\pm$0.006 & 1.756$\pm$0.222 & -0.004$\pm$0.278 & -0.018$\pm$0.254 &     DCC \\
 4 &     AR Aur & B9V + B9.6V &    172.768 &     -2.233 &    8.20&       0.10 &     7.89& 0.000$^{d}$ & 6.15$\pm$0.007 & -0.043$\pm$0.007 & 6.190$\pm$0.019 & -0.064$\pm$0.025 & -0.011$\pm$0.029 &     AAA \\
 5 &     WW Aur &  A4m + A5m &    181.724 &     10.519 &    11.86 &       0.09 &     11.47&  0.000$^{e}$ & 5.82$\pm$0.005 & 0.188$\pm$0.007 & 5.498$\pm$0.021 & -0.001$\pm$0.033 & 0.018$\pm$0.033 &       AAA \\
 6 &     ZZ Boo &  F3V + F3V &     31.820 &     75.482 &    8.88 &       0.09 &      8.60& 0.012$^{e}$ & 6.78$\pm$0.005 & 0.402$\pm$0.006 & 5.982$\pm$0.021 & 0.115$\pm$0.043 & 0.037$\pm$0.044 &        AAA \\
 7 &     SV Cam &    F9V + K4V &    131.572 &     26.523 &    11.77 &       0.09 &    11.37 & 0.015$^{a}$ & 9.30$\pm$0.019 & 0.703$\pm$0.023 & 7.872$\pm$0.023 & 0.366$\pm$0.051 & 0.130$\pm$0.051 &        AAA \\
 8 &     AR Cas &  B4V + A6V &    112.466 &     -2.659 &    5.67 &       0.10 &      5.44 & 0.054$^{f}$ & 4.89$\pm$0.003 & -0.122$\pm$0.003& 5.092$\pm$0.023 & -0.035$\pm$0.054 & -0.037$\pm$0.055 &     EAA \\
 9 &   YZ Cas & A1Vm + F2V &   122.549 &     12.121 &     11.24 &       0.05 &    11.13 & 0.070$^{g}$ & 5.64$\pm$0.003 & 0.078$\pm$0.004 & 5.585$\pm$0.019 & -0.059$\pm$0.042 & 0.042$\pm$0.043 &        AAA \\
 10 &   V636 Cen &    G0V + G7V &    316.742 &     10.619 &    15.36 &       0.07 &    15.03 & 0.027$^{h}$ & 8.67$\pm$0.015 & 0.650$\pm$0.018 & 7.474$\pm$0.030 & 0.348$\pm$0.040 & 0.066$\pm$0.034 &        AAA \\
11 &     EK Cep & A2V + G5Vp &    107.724 &     12.653 &    6.53 &       0.09 &      6.32 & 0.000$^{i}$ & 7.88$\pm$0.011 & 0.069$\pm$0.013 &  7.632$\pm$0.026 & 0.051$\pm$0.037 & 0.022$\pm$0.040 &        AAA \\
12 &     RS Cha &  A8V + A8V &    292.551 &    -21.632 &    10.23 &       0.04 &    10.15 &   0.031$^{a}$ & 6.05$\pm$0.015 & 0.229$\pm$0.021 & 5.994$\pm$0.030 & 0.117$\pm$0.048 & 0.025$\pm$0.051 &        AAA \\
13 &     RZ Cha &  F5V + F5V &    298.410 &    -20.324 &    5.43 &       0.12 &    5.12 &   0.004$^{h}$ & 8.05$\pm$0.011 & 0.460$\pm$0.014 & 7.131$\pm$0.030 & 0.190$\pm$0.047 & 0.037$\pm$0.052 &        AAA \\
14 &     WY Cnc & G5V + K9V &    199.471 &     39.307 &    11.76 &       0.15 &    10.65 &    0.011$^{a}$ & 9.49$\pm$0.034 & 0.727$\pm$0.042 & 7.992$\pm$0.023 & 0.403$\pm$0.029 & 0.108$\pm$0.028 &        AAA \\
15 &  $\alpha$ CrB &  A0V + G5V &     41.870 &     53.772 &  43.65 &       0.02 &   43.59 & 0.007$^{a}$ & 2.22$\pm$0.003 & 0.032$\pm$0.003 & 2.249$\pm$0.242 & -0.145$\pm$0.315 & 0.188$\pm$0.415 &      DCD \\
16 &  V1143 Cyg &  F5V + F5V &     87.251 &     15.595 &    25.12 &       0.02 &    25.07&   0.005$^{j}$ & 5.89$\pm$0.003 & 0.482$\pm$0.004 & 4.979$\pm$0.020 & 0.139$\pm$0.028 & 0.068$\pm$0.028 &       AAA \\
17 &  DE Dra &  B9V + G2V    &     96.483 &     14.364 &    8.62&       0.06 &      8.49 & 0.040$^{a}$ & 5.71$\pm$0.003 & -0.043$\pm$0.007 & 5.695$\pm$0.021 & -0.029$\pm$0.029 & 0.019$\pm$0.026 &       AAA \\
18 &     TX Her & A5V + F0V &     66.872 &     34.440 &    5.55 &       0.15 &      4.99 &  0.000$^{k}$ & 8.11$\pm$0.010 & 0.292$\pm$0.014 & 7.535$\pm$0.023 & 0.049$\pm$0.030 & 0.095$\pm$0.030 &        AAA \\
19 &   V624 Her & A3Vm + A7V:  &     38.716 &     21.250 &    6.93 &       0.11 &   6.60 &  0.051$^{l}$ & 6.18$\pm$0.005 & 0.211$\pm$0.006 & 5.740$\pm$0.019 & 0.034$\pm$0.026 & 0.056$\pm$0.018 &        AAF \\
20 &   V772 Her & G0V + G5V &     47.757 &     19.298 &    26.51 &       0.05 &     26.23 &   0.026$^{h}$ & 7.07$\pm$0.006 & 0.654$\pm$0.007 & 5.818$\pm$0.030 & 0.337$\pm$0.050 & 0.127$\pm$0.049 &        AAA \\
21 &     HS Hya &  F4V + F4V &    261.349 &     31.676 &    11.04 &       0.08 &     10.75&  0.024$^{a}$ & 8.08$\pm$0.012 & 0.466$\pm$0.014 & 7.206$\pm$0.021 & 0.172$\pm$0.040 & 0.077$\pm$0.038 &        AAA \\
22 &     KW Hya & A5Vm + F0V &    237.548 &     26.953 &    12.10 &       0.07 &     11.85 &  0.012$^{m}$ & 6.10$\pm$0.005 & 0.232$\pm$0.008 & 5.654$\pm$0.023 & 0.022$\pm$0.046 & 0.088$\pm$0.045 &        AAA \\
23 &     GZ Leo &    K1V + K1V &    217.241 &     64.826 &    18.43 &       0.06 &   18.11 &  0.007$^{a}$ & 8.96$\pm$0.019 & 0.872$\pm$0.026 & 7.244$\pm$0.019 & 0.428$\pm$0.025 & 0.111$\pm$0.023 &        AAA \\
24 &     UV Leo &  G0V + G2V &    228.700 &     56.462 &    10.85 &       0.11 &     10.33 &  0.016$^{a}$ & 8.91$\pm$0.019 & 0.657$\pm$0.023 & 8.069$\pm$0.026 & 0.318$\pm$0.033 & 0.108$\pm$0.026 &        AAA \\
25 &     UW LMi &  F8V + F8V &    202.219 &     61.777 &    7.73 &       0.14 &      7.07 & 0.003$^{c}$ & 8.34$\pm$0.012 & 0.596$\pm$0.015 & 7.320$\pm$0.034 & 0.281$\pm$0.043 & 0.059$\pm$0.031 &        AAA \\
26 &     GG Lup &  B7V + B9V &    330.846 &     13.954 &    6.34 &       0.11 &      5.99 & 0.027$^{n}$ & 5.59$\pm$0.005 & -0.099$\pm$0.007 &6.117$\pm$0.020 & -0.137$\pm$0.039 & 0.042$\pm$0.040 &      AAA \\
27 &     FL Lyr &  F8V + G8V &     77.268 &     15.933 &    7.69 &       0.12 &      7.25 & 0.049$^{o}$ & 9.35$\pm$0.022 & 0.574$\pm$0.025 & 8.243$\pm$0.026 & 0.258$\pm$0.042 & 0.086$\pm$0.039 &        AAA \\
28 &   V478 Lyr &  G8V + M2V &     61.852 &     10.124 &    35.70&       0.02 &      35.63 & 0.008$^{a}$ & 7.78$\pm$0.007 & 0.763$\pm$0.009 & 6.232$\pm$0.020 & 0.377$\pm$0.026 & 0.114$\pm$0.026 &        AAA \\
29 &     TZ Men &  B9V + G1V &    297.348 &    -28.827 &   9.35 &       0.05 &      9.24 & 0.055$^{a}$ & 6.18$\pm$0.004 & -0.003$\pm$0.004 & 6.137$\pm$0.029 & 0.018$\pm$0.049 & 0.002$\pm$0.048 &        AAA \\
30 &     UX Men &  F8V + F8V &    287.845 &    -31.096 &     9.93 &       0.06 &    9.77 &   0.027$^{p}$ & 8.23$\pm$0.010 & 0.549$\pm$0.011 & 7.195$\pm$0.027 & 0.219$\pm$0.038 & 0.061$\pm$0.036 &        AAA \\
31 & $\eta$ Mus &  B8V + B9V &    305.178 &    -5.127 &    8.04 &       0.07 &     7.86 &  0.008$^{q}$ & 4.79$\pm$0.002 & -0.078$\pm$0.003 & 4.949$\pm$0.044 & -0.082$\pm$0.050 & 0.040$\pm$0.033 &      EAA \\
32 &     EE Peg & A3mV + F5V &     64.182 &    -31.117 &    7.61 &       0.12 &     7.14 &  0.020$^{a}$ & 6.96$\pm$0.015 & 0.120$\pm$0.018 & 6.720$\pm$0.018 & 0.003$\pm$0.044 & 0.071$\pm$0.049 &        AAA \\
33 &   V505 Per &  F5V + F5V &    135.812 &     -6.101 &    15.00 &       0.06 &    14.81 &   0.012$^{a}$ & 6.86$\pm$0.007 & 0.456$\pm$0.001 & 6.070$\pm$0.067 & 0.277$\pm$0.076 & 0.022$\pm$0.041 &        AAA \\
34 &   V570 Per &  F5V + F5V &    145.178 &     -8.189 &    8.53 &       0.11 &     8.07 &  0.071$^{a}$ & 8.05$\pm$0.012 & 0.491$\pm$0.015 & 7.160$\pm$0.026 & 0.212$\pm$0.031 & 0.066$\pm$0.026 &        AAA \\
35 &    $\zeta$ Phe &  B6V + B8V &    297.833 &  -61.714 &  11.66 &       0.07 &  11.45 & 0.000$^{r}$ & 3.94$\pm$0.004 & -0.120$\pm$0.220 & 4.216$\pm$0.450 & 0.001$\pm$0.570 & 0.011$\pm$0.539 &       DDD \\
36 &     UV Psc &  G5V + K3V &    134.149 &    -55.504 &    15.87 &      0.08 &    15.42 &   0.017$^{a}$ & 8.98$\pm$0.023 & 0.712$\pm$0.028 & 7.633$\pm$0.029 & 0.407$\pm$0.040 & 0.092$\pm$0.035 &        AAA \\
37 &     BB Scl & K3V + K4V &    231.697 &    -80.036 &    42.29 &       0.03 &    42.08 &    0.003$^{a}$ & 7.11$\pm$0.009 & 0.909$\pm$0.011 & 5.340$\pm$0.023 & 0.367$\pm$0.079 & 0.283$\pm$0.078 &        AEA \\
38 &     CD Tau &  F6V + F6V &    183.955 &    -10.136 &    13.66 &      0.12 &    12.82 &   0.026$^{h}$ & 6.69$\pm$0.009 & 0.523$\pm$0.009 & 5.851$\pm$0.021 & 0.183$\pm$0.037 & 0.079$\pm$0.042 &        AAA \\
39 &   V818 Tau &    G6V + K6V &    177.621 &    -23.356 &    21.40 &    0.06 &    21.11 & 0.005$^{s}$ & 8.32$\pm$0.017 & 0.756$\pm$0.004 &  6.865$\pm$0.020 & 0.379$\pm$0.026 & 0.089$\pm$0.024 &        AAA \\
40 &  V1229 Tau &  A0Vp + Am &    166.535 &    -23.319 &   9.05 &      0.11 &      8.61 & 0.025$^{t}$ & 6.83$\pm$0.011 & 0.066$\pm$0.006 & 6.635$\pm$0.023 & -0.006$\pm$0.035 & 0.034$\pm$0.035 &        AAA \\
41 &     XY UMa & G9V + K7V &    162.720 &     41.675 &    15.09 &       0.10 &    14.49 &    0.005$^{a}$ & 9.50$\pm$0.029 & 0.765$\pm$0.039 & 7.770$\pm$0.020 & 0.493$\pm$0.043 & 0.118$\pm$0.042 &        AAA \\
42 &     PT Vel &    A1V + A6V &     266.180 &    3.329 &    6.20 &      0.10 &    5.94 & 0.004$^{u}$ & 7.02$\pm$0.006 & 0.055$\pm$0.008 & 6.863$\pm$0.030 & 0.010$\pm$0.042 & 0.016$\pm$0.041 &        AAA \\
43 &     ER Vul &    G0V + G5V &     73.342 &    -12.306 &    20.06 &    0.04 &   19.92 &  0.018$^{h}$ & 7.33$\pm$0.007 & 0.614$\pm$0.010 & 6.082$\pm$0.019 & 0.294$\pm$0.028 & 0.070$\pm$0.029 &        AAA \\
44 &     HD 71636 &    F2V + F5V &     184.980 &    34.803 &  8.54&     0.11 &    8.10 & 0.026$^{h}$ & 7.88$\pm$0.010 & 0.441$\pm$0.015 & 7.074$\pm$0.020 & 0.152$\pm$0.029 & 0.036$\pm$0.040&        AAA \\
\hline
\end{tabular}  
{
\\
(a) \cite{Schlegeletal1998}, (b) \cite{Popper1981}, (c) \cite{NordstromJohansen1994b}, (d) \cite{NordstromJohansen1994a}, (e) \cite{Lacy1979},  (f) \cite{Holmgrenetal1999}, (g) \cite{deLandtsheer1983}, (h) \cite{Nordstrometal2004}, (i) \cite{Popper1987}, (j) \cite{Andersenetal1987}, (k) \cite{Popper1980}, (l) \cite{Popper1984}, (m) \cite{Andersenvaz1984}, (n) \cite{Andersenetal1993}, (o) \cite{Lacy2002}, (p) \cite{Andersenetal1989}, (q) \cite{Bakisetal2007}, (r) \cite{Clausenetal1976}, (s) \cite{Lastennetetal1999}, (t) \cite{Groenewegenetal2007}, (u) \cite{Bakisetal2008} 
}
}
\end{table*}


\begin{table*}
\caption{Absolute magnitudes and distances calculated from LK corrected {\em Hipparcos} parallaxes 
and Eqs. 5-6 for 44 detached binaries.}
\begin{center}
\begin{tabular}{clcccccccc}
\hline
   &           &              Hip  &        Eq. 5    &    Hip          &    Eq. 6        &      Hip   &        Hip  &    Eq. 5  &   Eq. 6    \\
ID &  Star     &   $M_{V_{LK}}$   &  $M_{V}$    & $M_{J_{LK}}$    &   $M_{J}$  & $d$ (pc)   &$d_{LK}$ (pc)& $d$ (pc)& $d$ (pc)\\
\hline
1 &      RT And &  4.456$\pm$0.201 & 3.312$\pm$0.125 & 3.567$\pm$0.202 & 2.471$\pm$0.153 &  75$\pm$6  &  78$\pm$6  &  132$\pm$7 &  129$\pm$6 \\
2 &    V805 Aql &  0.785$\pm$0.336 & 1.126$\pm$0.113 & 0.443$\pm$0.353 & 1.135$\pm$0.204 & 172$\pm$26 & 192$\pm$26 &  164$\pm$8 &  139$\pm$8 \\
3 & $\beta$ Aur & -0.108$\pm$0.048 & 0.659$\pm$0.101 &       $-$       &      $-$        &  25$\pm$1  &  25$\pm$1  &   18$\pm$1 &       $-$  \\
4 &      AR Aur &  0.635$\pm$0.213 & -0.050$\pm$0.104 & 0.675$\pm$0.225 & 0.205$\pm$0.153 & 122$\pm$12 & 127$\pm$12 &  174$\pm$9 &  157$\pm$9 \\
5 &      WW Aur &  1.118$\pm$0.198 & 1.315$\pm$0.102 & 0.796$\pm$0.214 & 0.714$\pm$0.167 &  84$\pm$8  &  87$\pm$8  &   80$\pm$4 &   91$\pm$4 \\
6 &      ZZ Boo &  1.415$\pm$0.196 & 2.508$\pm$0.101 & 0.643$\pm$0.212 & 1.405$\pm$0.188 & 113$\pm$10 & 116$\pm$10 &   70$\pm$4 &   82$\pm$4 \\
7 &      SV Cam &  4.533$\pm$0.216 & 4.269$\pm$0.132 & 3.138$\pm$0.220  & 3.281$\pm$0.205 &  85$\pm$8  &  88$\pm$8  &   99$\pm$6 &   82$\pm$5 \\
8 &      AR Cas & -1.599$\pm$0.218 &-0.836$\pm$0.096 &        $-$      &        $-$      & 176$\pm$18 & 184$\pm$18 &  129$\pm$7 &       $-$  \\
9 &      YZ Cas &  0.655$\pm$0.109 & 0.251$\pm$0.097 & 0.755$\pm$0.125 & 0.359$\pm$0.184 &  89$\pm$4  &  90$\pm$4  &  108$\pm$6 &  108$\pm$5 \\
10 &   V636 Cen &  4.471$\pm$0.173 & 3.885$\pm$0.123 & 3.335$\pm$0.188 & 2.758$\pm$0.184 &  65$\pm$5  &  67$\pm$5  &   87$\pm$5 &   87$\pm$5 \\
11 &     EK Cep &  1.884$\pm$0.204 & 0.612$\pm$0.114 & 1.636$\pm$0.219 & 1.011$\pm$0.183 & 153$\pm$14 & 158$\pm$14 &  284$\pm$15&  211$\pm$12\\
12 &     RS Cha &  0.986$\pm$0.113 & 1.374$\pm$0.126 & 0.999$\pm$0.128 & 1.285$\pm$0.209 &  98$\pm$4  &  99$\pm$4  &   82$\pm$5 &   86$\pm$5 \\
13 &     RZ Cha &  1.584$\pm$0.263 & 2.898$\pm$0.115 & 0.673$\pm$0.282 & 1.819$\pm$0.209 & 184$\pm$22 & 195$\pm$22 &  107$\pm$5 &  115$\pm$7 \\
14 &     WY Cnc &  4.593$\pm$0.351 & 4.434$\pm$0.166 & 3.119$\pm$0.340 & 3.350$\pm$0.160 &  85$\pm$13 &  94$\pm$13 &  101$\pm$6 &   84$\pm$5 \\
15 &$\alpha$ CrB & 0.395$\pm$0.042 & 0.352$\pm$0.096 &        $-$      &        $-$      &  23$\pm$1  &  23$\pm$1  &   23$\pm$2 &       $-$  \\
16 &  V1143 Cyg &  2.871$\pm$0.051 & 3.022$\pm$0.097 & 1.971$\pm$0.068 & 1.739$\pm$0.156 &  40$\pm$1  &  40$\pm$1  &   37$\pm$2 &   44$\pm$3 \\
17 &     DE Dra &  0.231$\pm$0.133 &-0.286$\pm$0.100 & 0.305$\pm$0.151 & 0.463$\pm$0.156 &  116$\pm$7 & 118$\pm$7  &  149$\pm$8 &  109$\pm$6 \\
18 &     TX Her &  1.601$\pm$0.338 & 1.929$\pm$0.114 & 1.026$\pm$0.351 & 1.452$\pm$0.163 &  180$\pm$28& 200$\pm$28 &  172$\pm$9 &  165$\pm$8 \\
19 &   V624 Her &  0.120$\pm$0.237 & 1.149$\pm$0.101 &        $-$      &        $-$      &  144$\pm$16& 152$\pm$16 &   94$\pm$5 &       $-$  \\
20 &   V772 Her &  4.083$\pm$0.117 & 3.914$\pm$0.103 & 2.889$\pm$0.141 & 3.083$\pm$0.209 &   38$\pm$2 &  38$\pm$2  &   41$\pm$2 &   35$\pm$2 \\
21 &     HS Hya &  3.163$\pm$0.186 & 2.815$\pm$0.116 & 2.342$\pm$0.195 & 1.917$\pm$0.179 &   91$\pm$7 &  93$\pm$7  &  109$\pm$6 &  113$\pm$6 \\
22 &     KW Hya &  1.432$\pm$0.159 & 1.504$\pm$0.103 & 1.012$\pm$0.177 & 1.234$\pm$0.194 &   83$\pm$6 &  84$\pm$6  &   82$\pm$4 &   76$\pm$4 \\
23 &     GZ Leo &  5.228$\pm$0.160 & 5.314$\pm$0.135 & 3.528$\pm$0.160 & 3.515$\pm$0.147 &   54$\pm$4 &  55$\pm$4  &   53$\pm$3 &   56$\pm$2 \\
24 &     UV Leo &  3.931$\pm$0.251 & 3.991$\pm$0.132 & 3.126$\pm$0.258 & 2.894$\pm$0.165 &   92$\pm$10&  97$\pm$10 &   94$\pm$5 &  108$\pm$6 \\
25 &     UW LMi &  2.578$\pm$0.316 & 3.707$\pm$0.117 & 1.564$\pm$0.338 & 2.431$\pm$0.188 &  129$\pm$18& 141$\pm$18 &   84$\pm$5 &   95$\pm$5 \\
26 &     GG Lup & -0.607$\pm$0.252 &-0.540$\pm$0.102 &-0.020$\pm$0.267 & 0.074$\pm$0.179 &  158$\pm$18& 167$\pm$18 &  162$\pm$8 &  160$\pm$8 \\
27 &     FL Lyr &  3.500$\pm$0.274 & 3.306$\pm$0.137 & 2.502$\pm$0.278 & 2.349$\pm$0.187 &  130$\pm$15& 138$\pm$15 &  151$\pm$9 &  148$\pm$8 \\
28 &   V478 Lyr &  5.514$\pm$0.055 & 4.665$\pm$0.106 & 3.984$\pm$0.068 & 3.262$\pm$0.152 &   28$\pm$1 &  28$\pm$1  &   41$\pm$3 &   39$\pm$2 \\
29 &     TZ Men &  0.838$\pm$0.119 &-0.139$\pm$0.098 & 0.916$\pm$0.144 & 0.559$\pm$0.206 &  107$\pm$6 & 108$\pm$6  &  170$\pm$8 &  128$\pm$7 \\
30 &     UX Men &  3.095$\pm$0.145 & 3.288$\pm$0.111 & 2.120$\pm$0.162 & 2.052$\pm$0.181 &  101$\pm$6 & 102$\pm$6  &   94$\pm$5 &  106$\pm$5 \\
31 & $\eta$ Mus & -0.758$\pm$0.160 &-0.304$\pm$0.095 &       $-$       &        $-$      &  124$\pm$9 & 127$\pm$9  &  103$\pm$5 &       $-$  \\
32 &     EE Peg &  1.166$\pm$0.275 & 0.795$\pm$0.123 & 0.970$\pm$0.278 & 1.007$\pm$0.191 &  131$\pm$16& 140$\pm$16 &  166$\pm$9 &  138$\pm$7 \\
33 &   V505 Per &  2.676$\pm$0.123 & 2.827$\pm$0.092 & 1.912$\pm$0.189 & 2.159$\pm$0.264 &   67$\pm$4 &  68$\pm$4  &   63$\pm$3 &   60$\pm$5 \\
34 &   V570 Per &  2.364$\pm$0.257 & 2.685$\pm$0.117 & 1.631$\pm$0.271 & 1.924$\pm$0.163 &  117$\pm$13& 124$\pm$13 &  107$\pm$6 &  108$\pm$6 \\
35 &$\zeta$ Phe & -0.766$\pm$0.147 &-0.505$\pm$0.314 &        $-$      &        $-$      &   86$\pm$6 &  87$\pm$6  &   77$\pm$4 &        $-$ \\
36 &     UV Psc &  4.867$\pm$0.203 & 4.310$\pm$0.141 & 3.558$\pm$0.209 & 3.260$\pm$0.184 &   63$\pm$5 &  65$\pm$5  &   84$\pm$5 &   74$\pm$5 \\
37 &     BB Scl &  5.221$\pm$0.085 & 5.557$\pm$0.110 &        $-$      &        $-$      &   24$\pm$1 &  24$\pm$1  &   20$\pm$1 &       $-$  \\
38 &     CD Tau &  2.148$\pm$0.268 & 3.140$\pm$0.107 & 1.367$\pm$0.281 & 1.981$\pm$0.180 &   73$\pm$9 &  78$\pm$9  &   49$\pm$3 &   59$\pm$3 \\
39 &   V818 Tau &  4.927$\pm$0.143 & 4.641$\pm$0.111 & 3.483$\pm$0.146 & 3.123$\pm$0.150 &   47$\pm$3 &  47$\pm$3  &   54$\pm$3 &   56$\pm$3 \\
40 &  V1229 Tau &  1.428$\pm$0.243 & 0.446$\pm$0.107 & 1.288$\pm$0.255 & 0.714$\pm$0.173 &  110$\pm$12& 116$\pm$12 &  183$\pm$9 &  151$\pm$8 \\
41 &     XY UMa &  5.290$\pm$0.242 & 4.694$\pm$0.158 & 3.571$\pm$0.233 & 3.899$\pm$0.185 &   66$\pm$7 &  69$\pm$7  &   91$\pm$5 &   59$\pm$3 \\
42 &     PT Vel &  0.877$\pm$0.223 & 0.505$\pm$0.104 & 0.728$\pm$0.247 & 0.748$\pm$0.193 &  161$\pm$16& 168$\pm$16 & 200$\pm$10 &  167$\pm$9 \\
43 &     ER Vul &  3.770$\pm$0.098 & 3.725$\pm$0.107 & 2.562$\pm$0.110 & 2.528$\pm$0.156 &   50$\pm$2 &  50$\pm$2  &   51$\pm$3 &   51$\pm$3 \\
44 &   HD 71636 &  2.341$\pm$0.249 & 2.656$\pm$0.115 & 1.593$\pm$0.259 & 1.553$\pm$0.169 &  117$\pm$13& 123$\pm$13 &  107$\pm$5 &  126$\pm$6 \\
\hline
\end{tabular}
\end{center}
\end{table*}

\section{Absolute magnitude calibrations}
Selected colours and LK corrected absolute magnitudes $M_{V}$ and $M_{J}$ according to 
{\em Hipparcos} distances for the sample stars (Tables 1 and 2) were used in calibrating 
the coefficients of the following equations:  

\begin{equation}
M_{V}=a(B-V)_{o}+c, \\
\end{equation}

\begin{equation}
M_{J}=a(J-H)_{o}+b(H-K_{s})_{o}+c. \\
\end{equation}
The coefficients of each equation and their 1$\sigma$ errors have been determined by 
the method of least squares. All 44 systems in Table 1 were used to predict the 
coefficients for the relation using optical colour (Eq. 5). However, only 37 out of 
44 were used in the calibrations involving  {\em 2MASS} data since observed colours 
with the quality flag ``AAA'' were preferred for them. The predicted coefficients 
and their uncertainties together with the correlation coefficients, standard 
deviations and standard errors are given in Table 3. Consequently, it can be concluded 
that each equation above would provide an absolute magnitude with an internal uncertainty of 
$\pm0.09$ mag. Moreover, since the correlation coefficients are 0.91 and 0.95 (Table 3), 
strong correlations between the involved parameters (absolute magnitude and colours) 
are confirmed. The calibrated equations can be considered valid within the ranges 
$-0.18<(B-V)_{o}<0.91$ and $-1.6<M_{V}<5.5$ for the calibration including optical 
colour, and $-0.15<(J-H)_{o}<0.50$, $-0.02<(H-K_{s})_{o}<0.13$ and $0<M_{J}<4$ 
for the {\em 2MASS} photometric system. These limits correspond to the spectral 
classes $B4-K7$ \citep{Cox2000,Coveyetal2007}. The calibrations are based on 
the trigonometric parallaxes with the relative errors $(\sigma_{\pi}/\pi)\leq0.15$.

\begin{figure}
\center
\resizebox{80mm}{75.5mm}{\includegraphics*{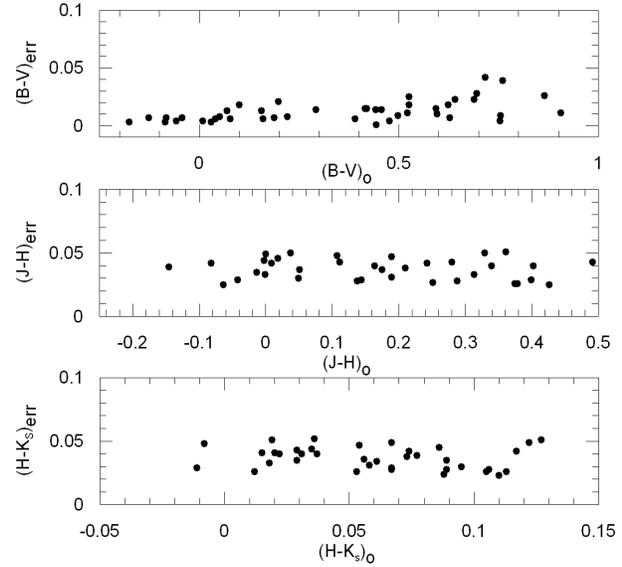}} 
\caption{Colour errors for the optical and {\em 2MASS} data.}
\end{figure}

\begin{figure}
\center
\includegraphics[scale=0.6, angle=0]{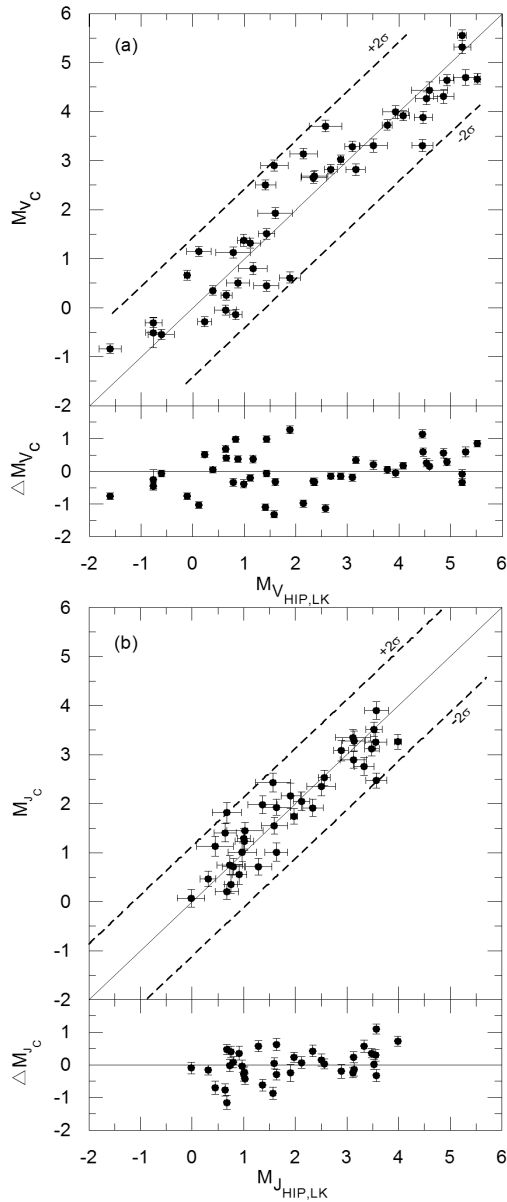}
\caption{Absolute magnitudes predicted by the Eqs. 5 (a) and 6 (b) 
versus visual and near-infrared absolute magnitudes calculated from 
LK corrected {\em Hipparcos} parallaxes.}
\end{figure} 

\begin{figure}
\center
\includegraphics[scale=0.6, angle=0]{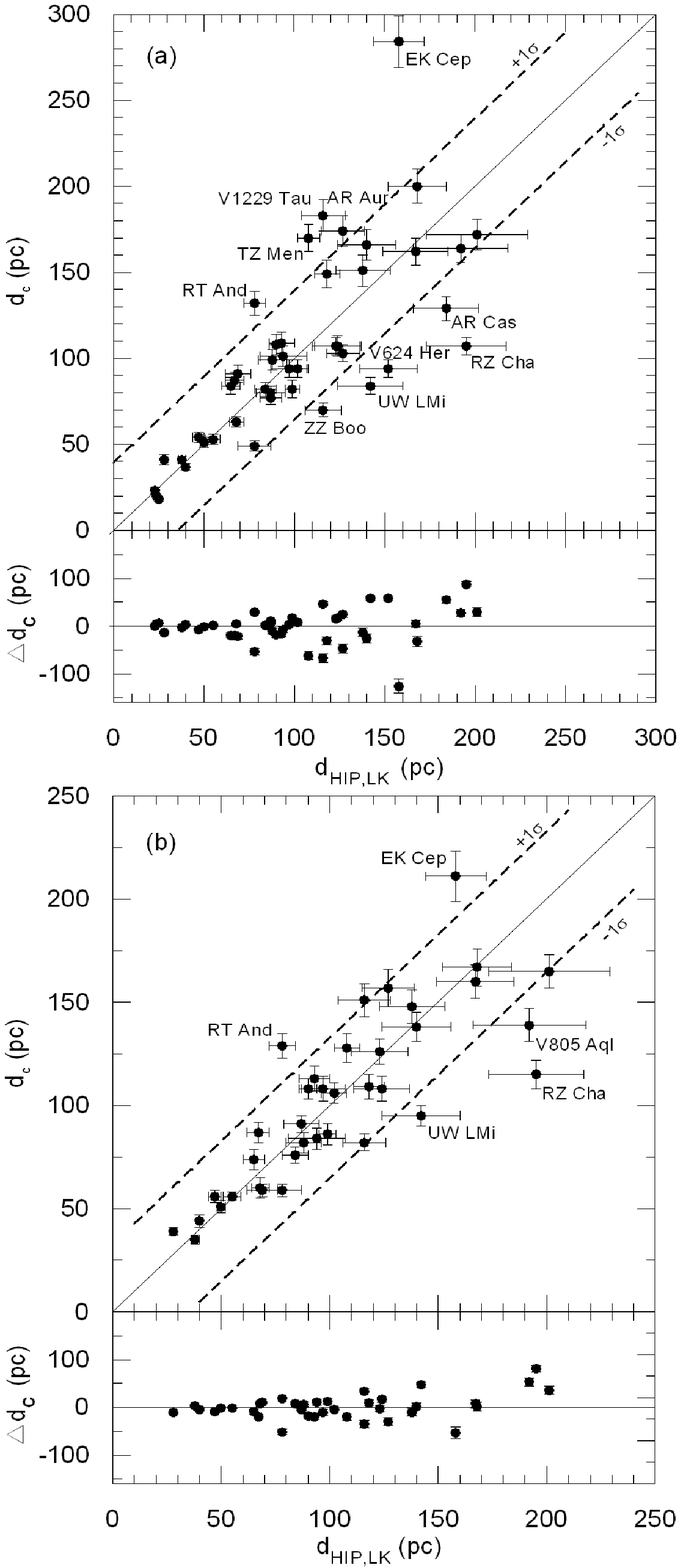}
\caption{Comparison of the distances calculated from LK corrected {\em Hipparcos} 
parallaxes with the predicted distances from Eqs. 5(a) and 6 (b).}
\end{figure} 

\subsection{Uncertainties of the calibrations}
In order to see how accurate the observed colours are, the errors of the observed 
colours have been plotted against the intrinsic colours in Fig. 1 where the 
visual colour $((B-V)_{0})$ have smaller uncertainties than the near-infrared 
colours $((J-H)_{0}$, $(H-K_{s})_{0})$. The mean observational errors are about 
0.02 ($\sigma=\pm0.02$) and 0.04 ($\sigma=\pm0.01$) mag in the optical and 
near-infrared colours, respectively. This is not surprising, because {\em 2MASS} 
magnitudes were obtained from single-epoch observations, while optical 
observations have been done more than once. These mean errors of colours 
introduce typically $\pm$0.17 and $\pm$0.22 mag uncertainties at $M_{V}$ and 
$M_{J}$, respectively.

Uncertainty in the de-reddening introduces an additional uncertainty. Due to smaller 
de-reddening at infrared colours, infrared calibrations are expected to be less 
affected by de-reddening process.   
 
Some of the binaries can be unrecognized as triple or multiple systems. Thus, 
their parallax measurements cannot be reliable. Assuming that the parallaxes of 
systems in our sample are precise enough, a median ($\sigma_{\pi}/\pi$) 
value of 0.07 of the sample introduces a typical error of $\pm$0.15 mag. 

\subsection{Comparison with {\em Hipparcos} absolute magnitudes and distances}

Absolute magnitudes predicted by LCs relations are compared to the absolute 
magnitudes obtained from {\em Hipparcos} catalogue for the calibration stars 
(Fig. 2). Almost all calibration stars are located in the prediction limits 
of 2$\sigma$. Fig. 2 and Table 3 show that absolute magnitudes calculated 
from the calibration equation using near-infrared colours have stronger 
correlation and less scatter on diagonal.

The distances predicted from the distance modulus have been compared in a similar 
manner in Fig. 3. Standard deviations of the residuals from the {\em Hipparcos} 
distances are 37 and 26 pc for optical and near infrared colours, respectively. 
These values show that the calibration equation including near-infrared colours 
are better in correlation as expected than the absolute magnitude 
comparisons given above. In Fig. 3a-b, we showed the 1$\sigma$ prediction limits. 
Systems located out of 1$\sigma$ prediction limits are indicated. It is 
interesting to note that these scattered systems are almost the same as in 
Fig. 3a-b.

\begin{table}
\setlength{\tabcolsep}{1.3pt}
\small{
\caption{Least square coefficients of Eqs. 5-6. $R$, $s$ and $se$ denote correlation 
coefficient, standard deviation and standard error, respectively.}
\begin{tabular}{ccccccc}
\hline
Eq. &        $a$ &          $b$        &          $c$            &  $R$ &  $s$ & $se$ \\
\hline
5 & 5.908($\pm$0.309) &              $-$  & 0.204($\pm$0.143) & 0.95 & 0.63 & 0.09 \\
6 & 5.228($\pm$0.715) & 6.185($\pm$3.173) & 0.608($\pm$0.154) & 0.91 & 0.49 & 0.08 \\
\hline
\end{tabular}
}  
\end{table}

\begin{table*}
\setlength{\tabcolsep}{2pt}
\caption{Detached systems with main sequence components and known distances 
collected from the literature. The last column of the table gives references 
including information for colour excess, distance given in literature and remarks, 
in order.}
{\small
\begin{tabular}{clccccccccccccl}
\hline
  &      &        &        &       &         &               &      &     Eq. 5 &   Eq. 6 &  Eq. 5 &  Eq. 6 & Literature &     & \\
ID &  Star & $V$ & $(B-V)$ & $J$ & $(J-H)$& $(H-K_{s})$& $E(B-V)$ & $M_{V}$ & $M_{J}$&  $d$ (pc)&  $d$ (pc)& $d$ (pc)& Rem$^{*}$ &    Refs$^{**}$ \\
\hline
 1 &     KZ And &  7.91 &  0.900 & 6.225 &      0.437 &      0.129 &      0.010 &      5.462 &      3.662 &   30$\pm$2 &   32$\pm$3 &   25$\pm$5 &            & (01, 02, -) \\
 2 &     KP Aql &  9.40 &  0.446 & 8.521 &      0.067 &      0.050 &      0.130 &      2.071 &      0.900 & 243$\pm$12 & 317$\pm$26 & 282$\pm$12 &            & (03, 03, -) \\
 3 &   V432 Aur &  8.01 &  0.566 & 6.864 &      0.277 &      0.069 &      0.000 &      3.548 &      2.483 &   78$\pm$5 &   75$\pm$6 & 118$\pm$23 &            & (04, 02, -) \\
 4 &     AD Boo &  9.45 &  0.521 & 8.389 &      0.232 &      0.051 &      0.000 &      3.282 &      2.136 &  171$\pm$9 & 178$\pm$15 &  203$\pm$8 &            & (03, 03, -) \\
 5 &     CV Boo & 10.75 &  0.726 & 9.570 &      0.288 &      0.081 &      0.014 &      4.410 &      2.570 & 182$\pm$11 & 250$\pm$20 & 235$\pm$24 &            & (01, 05, -) \\
 6 &     AS Cam &  8.59 & -0.008 & 8.524 &     -0.022 &      0.007 &      0.080 &     -0.316 &      0.308 & 539$\pm$28 & 426$\pm$42 & 480$\pm$48 &       T, A & (06, 07, 08) \\
 7 &     WW Cam & 10.16 &  0.510 & 9.182 &      0.065 &      0.081 &      0.397 &      0.872 &      0.334 & 409$\pm$21 & 500$\pm$39 & 376$\pm$12 &          A & (09,  09, 10)\\
 8 &   V392 Car &  9.48 &  0.191 & 9.053 &      0.080 &      0.037 &      0.101 &      0.736 &      0.971 & 485$\pm$28 & 397$\pm$34 & 349$\pm$28 &            & (11, 11, -) \\
 9 &     IT Cas & 11.15 &  0.488 &10.212 &      0.255 &      0.042 &      0.062 &      2.721 &      2.028 & 444$\pm$13 & 422$\pm$34 & 504$\pm$26 &          A & (12,12,10) \\
10 &     PV Cas &  9.75 &  0.171 & 9.324 &     -0.017 &      0.046 &      0.217 &     -0.068 &      0.196 & 675$\pm$34 & 613$\pm$51 & 670$\pm$67 &          A & (13, 13, 25) \\
11 &     SZ Cen &  8.52 &  0.305 & 8.346 &      0.127 &      0.065 &      0.060 &      1.651 &      1.507 & 217$\pm$13 & 228$\pm$22 & 262$\pm$101&            & (14, 02, -) \\
12 &     EY Cep &  9.80 &  0.367 & 9.035 &      0.058 &      0.136 &      0.049 &      2.083 &      1.613 & 326$\pm$17 & 299$\pm$29 & 307$\pm$14 &          A & (15, 15, 10) \\
13 &     ZZ Cep &  8.63 &  0.285 & 8.148 &      0.124 &      0.021 &      0.148 &      1.013 &      0.968 & 270$\pm$14 & 257$\pm$30 & 168$\pm$89 &            & (01, 02, -) \\
14 &     TV Cet &  8.61 &  0.439 & 7.753 &      0.131 &      0.073 &      0.067 &      2.402 &      1.555 & 158$\pm$10& 169$\pm$16 &  160$\pm$7  &       T, A & (01, 16, 08) \\
15 &     XY Cet &  8.74 &  0.280 & 8.212 &      0.018 &      0.079 &      0.100 &      1.267 &      0.912 & 271$\pm$19 & 277$\pm$25 & 212$\pm$76 &            & (17, 02, -) \\
16 &     GZ CMa &  7.97 &  0.159 & 7.613 &      0.002 &      0.052 &      0.063 &      0.771 &      0.761 & 251$\pm$15 & 229$\pm$21 & 270$\pm$86 &            & (18, 02, -) \\
17 &     MY Cyg &  8.31 &  0.377 & 7.692 &      0.115 &      0.055 &      0.055 &      2.106 &      1.393 & 161$\pm$9 & 178$\pm$21 & 264$\pm$64 &          A & (19, 02, 10) \\
18 &   V442 Cyg &  9.71 &  0.496 & 9.154 &      0.174 &      0.032 &      0.080 &      2.662 &      1.487 & 229$\pm$11 & 331$\pm$25 & 317$\pm$18 &            & (20, 03, -) \\
19 &   V477 Cyg &  8.51 &  0.168 & 8.099 &      0.023 &      0.017 &      0.040 &      0.960 &      0.722 & 306$\pm$18 & 294$\pm$23 & 205$\pm$15 &       T, A & (21, 21, 08) \\
20 &   V909 Cyg &  9.56 &  0.141 & 9.180 &      0.017 &      0.064 &      0.066 &      0.647 &      0.909 & 552$\pm$27 & 439$\pm$35 & 447$\pm$28 &        T   & (03, 03, 41) \\
21 &  V1061 Cyg &  9.24 &  0.592 & 8.122 &      0.205 &      0.086 &      0.296 &      1.953 &      1.381 & 188$\pm$11 & 198$\pm$16 &  167$\pm$6 &        T   & (01, 22, 22) \\
22 &     MR Del &  8.77 &  0.696 & 7.033 &      0.481 &      0.143 &      0.013 &      4.239 &      3.974 &   79$\pm$5 &   41$\pm$3 &  44$\pm$11 &        T   & (01, 02, 42) \\
23 &     BS Dra &  9.13 &  0.443 & 8.268 &      0.198 &      0.043 &      0.102 &      2.219 &      1.619 & 208$\pm$14 & 205$\pm$17 & 208$\pm$33 &            & (01, 02, -) \\
24 &     UZ Dra &  9.59 &  0.512 & 8.616 &      0.190 &      0.054 &      0.024 &      3.087 &      1.869 & 193$\pm$11 & 221$\pm$18 & 185$\pm$12 &            & (03, 03, -) \\
25 &     CW Eri &  8.40 &  0.389 & 7.799 &      0.140 &      0.033 &      0.011 &      2.437 &      1.511 &  153$\pm$9 & 180$\pm$17 & 168$\pm$37 &            & (23, 02, -) \\
26 &     RX Her &  7.26 &  0.054 & 7.256 &      0.015 &     -0.007 &      0.042 &      0.275 &      0.520 & 235$\pm$13 & 219$\pm$20 & 230$\pm$56 &          A & (24, 02, 25) \\
27 &     VZ Hya &  8.98 &  0.455 & 8.076 &      0.224 &      0.068 &      0.016 &      2.798 &      2.155 & 168$\pm$11 & 152$\pm$13 & 199$\pm$57 &            & (01, 02, -) \\
28 &     CM Lac &  8.20 &  0.190 & 7.843 &      0.018 &      0.057 &      0.060 &      0.972 &      0.887 & 256$\pm$14 & 240$\pm$27 & 227$\pm$45 &            & (26, 02, -) \\
29 &     RW Lac & 10.63 &  0.661 & 9.299 &      0.338 &      0.090 &      0.068 &      3.707 &      2.742 & 220$\pm$10& 199$\pm$16  & 190$\pm$10 &       A & (27, 27, 10) \\
30 &     TX Leo &  5.67 &  0.059 & 5.515 &      0.012 &      0.061 &      0.016 &      0.458 &      1.003 &  108$\pm$5 &   79$\pm$8 & 142$\pm$20 &          & (01, 02, -) \\
31 &     FS Mon &  9.60 &  0.388 & 8.826 &      0.175 &     -0.005 &      0.039 &      2.266 &      1.381 & 277$\pm$12 & 303$\pm$30 & 321$\pm$14 &            & (03, 03, -) \\
32 &     WZ Oph &  9.10 &  0.551 & 8.574 &      0.205 &      0.100 &      0.032 &      3.270 &      2.209 &  140$\pm$8 & 185$\pm$18 & 125$\pm$22 &            & (01, 02, -) \\
33 &   V451 Oph &  7.89 &  0.050 & 7.626 &     -0.011 &      0.064 &      0.155 &     -0.416 &      0.512 & 367$\pm$20 & 249$\pm$29 & 300$\pm$30 &          A & (28, 28, 25) \\
34 &     EW Ori &  9.94 &  0.612 & 8.808 &      0.204 &      0.106 &      0.014 &      3.737 &      2.285 & 171$\pm$11 & 201$\pm$24 &  180$\pm$5 &          A & (29, 29, 10) \\
35 &     GG Ori & 10.37 &  0.511 & 9.465 &      0.142 &      0.111 &      0.411 &      0.795 &      0.883 & 457$\pm$24 & 440$\pm$37 & 430$\pm$15 &          A & (01, 03, 10) \\
36 &     BK Peg &  9.99 &  0.540 & 8.892 &      0.249 &      0.032 &      0.050 &      3.099 &      1.968 & 222$\pm$11 & 238$\pm$19 & 260$\pm$26 &            & (30, 30, -) \\
37 &     OO Peg &  8.26 &  0.270 & 7.676 &      0.043 &      0.078 &      0.070 &      1.386 &      1.115 & 214$\pm$14 & 199$\pm$18 & 295$\pm$17 &            & (01, 31, -) \\
38 &     IQ Per &  7.73 &  0.041 & 7.561 &     -0.030 &      0.047 &      0.140 &     -0.381 &      0.352 & 343$\pm$19 & 261$\pm$25 & 275$\pm$15 &          A & (32, 32, 25) \\
39 &     PV Pup &  7.03 &  0.345 & 6.574 &      0.155 &      0.015 &      0.010 &      2.183 &      1.483 &  92$\pm$5 & 104$\pm$9  &   83$\pm$5 &          A &(01, 33, 10) \\
40 &     VV Pyx &  6.57 &  0.043 & 6.403 &     -0.007 &      0.036 &      0.022 &      0.328 &      0.733 &  172$\pm$9 & 135$\pm$11 & 195$\pm$10 &       A &(34, 34, 25) \\
41 &     AL Scl &  6.09 & -0.093 & 6.253 &     -0.055 &      0.018 &      0.027 &     -0.505 &      0.354 & 201$\pm$10 & 150$\pm$13 & 200$\pm$27 &            & (35, 35, -) \\
42 &     QX Ser &  8.68 &  0.635 & 7.477 &      0.266 &      0.098 &      0.054 &      3.637 &      2.454 &   94$\pm$5 &  99$\pm$9  &  95$\pm$10 &            & (23, 23, -) \\
43 &   V526 Sgr &  9.84 &  0.179 & 9.539 &      0.057 &      0.040 &      0.140 &      0.434 &      0.764 & 623$\pm$33 & 537$\pm$47 & 540$\pm$23 &          A & (36, 36, 25) \\
44 &  V1647 Sgr &  6.92 &  0.125 & 6.912 &     -0.016 &      0.041 &      0.041 &      0.706 &      0.667 &  165$\pm$9 & 175$\pm$18 & 160$\pm$10 &          A & (37, 37, 25) \\
45 &     ZZ UMa &  9.84 &  0.586 & 8.713 &      0.301 &      0.078 &      0.008 &      3.619 &      2.642 & 173$\pm$9 & 163$\pm$14  & 172$\pm$21 &            & (38, 38, -) \\
46 &     BH Vir &  9.65 &  0.650 & 8.459 &      0.274 &      0.113 &      0.035 &      3.837 &      2.645 &  138$\pm$10& 143$\pm$18 & 126$\pm$25 &            & (39, 02, -) \\
47 &     DM Vir &  8.74 &  0.469 & 7.792 &      0.181 &      0.083 &      0.023 &      2.839 &      2.006 & 147$\pm$9 & 142$\pm$16 &  192$\pm$4 &            & (40, 40, -) \\
48 &  Hip 56132 &  9.87 &  0.669 & 8.650 &      0.345 &      0.074 &      0.011 &      4.091 &      2.836 &  141$\pm$5 & 145$\pm$11 &  92$\pm$15 &            & (01, 02, -) \\
\hline
\end{tabular}
{
\\
($^{*}$) (T) Triple, (A) Apsidal motion\\
($^{**}$) (1) \cite{Schlegeletal1998}, (2) \cite{ESA97}, (3) \cite{Lacy2002}, (4) \cite{Siviero2004}, (5) \cite{Nelson2004}, (6) \cite{Maloney1989}, (7) \cite{Khodykin1997}, (8) \cite{Bozkurt2007}, (9) \cite{Lacyetal2002}, (10) \cite{Bulut2007}, (11) \cite{Debernardi2001}, (12) \cite{Lacyetal1997}, (13) \cite{Barembaum1995}, (14) \cite{Andersen1975}, (15) \cite{Lacy2006}, (16) \cite{Jorgensen1979}, (17) \cite{Srivastava1987}, (18) \cite{Popper1985}, (19) \cite{PopperEtzel1981}, (20) \cite{LacyFrueh1987}, (21) \cite{Degirmencietal2003}, (22) \cite{Torres2006}, (23) \cite{Nordstrometal2004}, (24) \cite{Lacy1979}, (25) \cite{Petrova1999}, (26) \cite{Popper1968}, (27) \cite{Lacyetal2005}, (28) \cite{Clausen1986}, (29) \cite{Popperetal1986}, (30) \cite{Demircan1994}, (31) \cite{Munari2001}, (32) \cite{Lacy1985}, (33) \cite{Vaz1984}, (34) \cite{Andersenetal1984}, (35) \cite{Haefneretal987}, (36) \cite{Lacy1997a}, (37) \cite{AndersenandGimenez1985}, (38) \cite{Clementetal1997a}, (39) \cite{Clementetal1997b}, (40) \cite{Latham1996}, (41) \cite{Lacy1997b}, (42) \cite{Cutispotoetal1997}
}
}
\end{table*}

\section{Application to the other detached systems}

In order to compare the distances estimated from our calibrations and various methods, 
48 detached systems other than our sample with main-sequence components and known 
distances were collected from the literature. Among these stars, distances of 16 
systems were taken from the {\em Hipparcos} catalogue with $0.15<(\sigma_{\pi}/\pi)\leq0.5$. 
For the rest of stars, distances were found from photometric analysis. 
Colours, colour excesses, absolute magnitudes and distances of the selected 
systems are listed in Table 4. Optical and near-infrared colours of these stars 
were taken from {\em Hipparcos} catalogue \citep{ESA97} and {\em 2MASS} Point-Source 
Catalogue \citep{Cutrietal2003}, respectively. $E(B-V)$ values were either taken 
from the literature or calculated from \cite{Schlegeletal1998} maps (see Section 2.1). 

The $M_{V}$ and $M_{J}$ absolute magnitudes of the binaries in Table 4 (columns 9-10) were 
estimated from Eqs. 5 and 6, respectively. These absolute magnitudes were 
used to calculate the distances by the Pogson's relation and recorded in next columns. 
The systems with known hypothetical third body or apsidal motion are indicated in 
column 14, as well.

The distances predicted from our LCs relations are compared to the distances used 
in literature, where various methods including trigonometric parallaxes were preferred 
(Fig. 4). At first glance, it looks like that predicted and collected distances are 
very much in agreement. Standard deviations of the residual distances are $\sim$50 pc 
for both the calibration equations using optical and near infrared colours. The 
systems with distances higher than $\sim$400 pc have the largest scatter. However, 
an eye inspection shows that scatter even for these distant systems is not higher 
than $\sim$100 pc, indicating a precision better than $\sim$ 25\%. In fact, 
the scatter in Fig. 4 are mostly due to uncertainties in distance measurements in 
the literature (particularly those from uncertain {\em Hipparcos} parallaxes), so 
the scatter in the calibrations should be smaller than the overall scatter.

\begin{figure}
\center
\resizebox{80mm}{145.3mm}{\includegraphics*{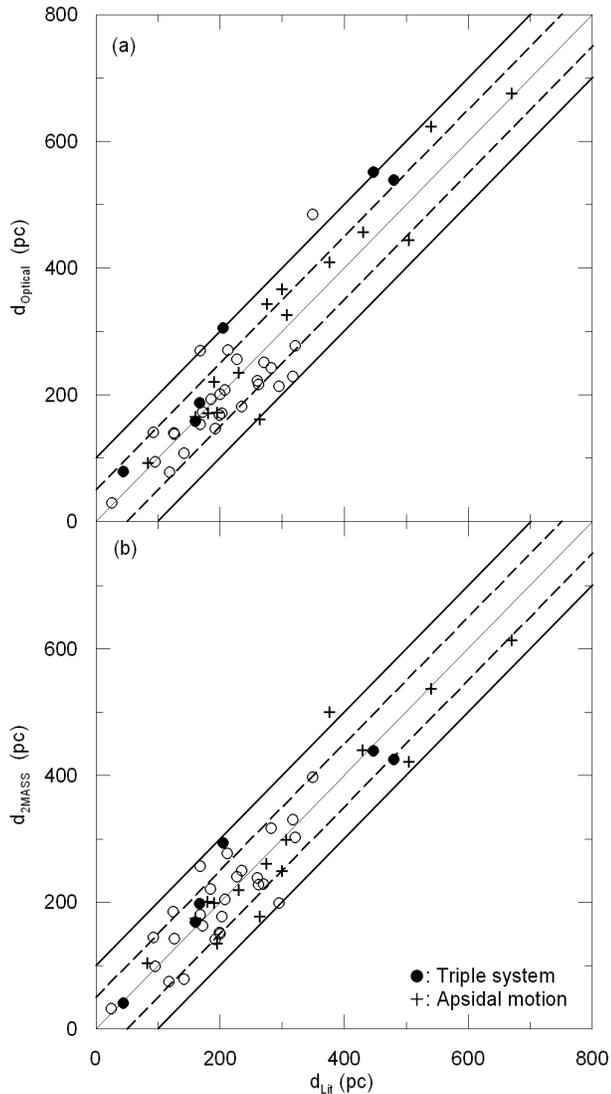}} 
\caption{The distances predicted from our LCs relations are compared to 
the distances used in literature, where various methods including trigonometric 
parallaxes were preferred. Stars with hypothetical third body or apsidal motion 
are indicated by different symbols. Vertical scale of (a) uses Eq. 5, while vertical 
axis of (b) uses Eq. 6. Thick-dashed and solid lines show 50 and 100 pc distances 
from the diagonal thin line representing equal values. Error bars are not 
shown for clarity.}
\end{figure} 

\section{Discussions and conclusions}

We have suggested absolute magnitude calibrations for detached binaries based on LK 
corrected {\em Hipparcos} parallaxes. The calibration equations cover a wide range 
of colours and absolute magnitudes. 

The distances of calibration stars were calculated using the absolute magnitudes 
obtained from the LCs relations and compared with the {\em Hipparcos} 
distances (Fig. 3). Although distances are in agreement in general (concentrated 
on diagonal), there is a considerable scatter for some systems. Large deviations 
from the LCs relations are mostly seen for the systems whose distances are higher 
than $\sim$100 pc with the exception of RT And, which is closer than 100 pc. 
We have listed these scattered systems in Table 5. The table shows that distances 
from literature agree better with our distance estimates from the calibration 
relation including near-infrared colours. The table also indicates that 
{\em Hipparcos} distances are not in good agreement with the distances recorded in 
literature, in general, although their relative parallax errors are smaller than 
0.15. One of the best examples is V1229 Tau. \cite{Southworthetal2005b} estimated 
the distance of V1229 Tau using three independent methods and found that 
{\em Hipparcos} distance is not in agreement with their results. Our estimates 
support \cite{Southworthetal2005b}, but not in agreement with {\em Hipparcos} 
distance. Thus, it seems that, in selecting stars for such calibration studies, 
not only relative parallax errors but also parallaxes higher than 10 mas must be 
carefully taken into account.

\begin{table}
\setlength{\tabcolsep}{4pt}
\center
\caption{The distances of scattered stars in calibration sample obtained from 
various other sources in literature. Columns 2, 3 and 4 include distances 
calculated using calibration Eqs. 5, 6 and LK corrected {\em Hipparcos} 
parallaxes, respectively. Column 5 shows the distances collected from literature.}
\begin{tabular}{lccccc}
\hline
           &      Eq. 5   &   Eq. 6   & Hip   &        Lit.  &     \\
      Star &     $d$ (pc) &     $d$ (pc) &  $d_{LK}$ (pc) &     $d$ (pc) & Refs\\
\hline
    RT And &    132$\pm$7 &    129$\pm$6 &    78$\pm$6  & 103          & (1) \\
  V805 Aql &    164$\pm$8 &    139$\pm$8 &   192$\pm$26 & --           & (-) \\
    AR Aur &    174$\pm$9 &    157$\pm$9 &   127$\pm$12 & 136$\pm$7    & (2) \\
    ZZ Boo &     70$\pm$4 &     82$\pm$4 &   116$\pm$10 & 79           & (3) \\
    AR Cas &    129$\pm$7 &           -- &   184$\pm$18 & --           & (-) \\ 
    EK Cep &   284$\pm$15 &    211$\pm$12&   158$\pm$14 & 190          & (4) \\
    RZ Cha &    107$\pm$5 &    115$\pm$7 &   195$\pm$22 & 190$\pm$10   & (5) \\
  V624 Her &     94$\pm$5 &           -- &   152$\pm$16 & 111          & (6) \\
    UW LMi &     84$\pm$5 &     95$\pm$5 &   141$\pm$18 &  82          & (7) \\
    TZ Men &    170$\pm$8 &    128$\pm$7 &   108$\pm$6  & 120$\pm$10   & (8) \\
 V1229 Tau &    183$\pm$9 &    151$\pm$8 &   116$\pm$12 & 139$\pm$4    & (9) \\
\hline
\end{tabular}
\\ 
{
(1) \cite{Arevaloetal1995}, (2) \cite{Semeniuk2000}, (3) \cite{Popper1998}, 
(4) \cite{Popper1987}, (5) \cite{Andersenetal1975}, (6) \cite{Lacy1979}, 
(7) \cite{Nordstrometal2004}, (8) \cite{Andersenetal1987b}, 
(9) \cite{Southworthetal2005b}  
}
\end{table}

It has been found that the distances indicated by the LCs relations and various 
methods are in agreement (see Fig. 4). Although there is some scattering in the 
figure, almost all systems are located very near to the diagonal line representing 
equal values. It is interesting to note that the systems in Fig. 4 with third body 
or apsidal motion do not display larger scatter in comparison to other stars. 

Normally, colour-magnitude relation exists for main- sequence single stars which 
is the main part of HR diagram itself. Similar colour-magnitude relation obviously 
exists also for main-sequence binaries with a deviation of about 0.50 mag 
(see Table 3). This could be explained by an argument that the greatest 
deviation from the relation of single star must occur when two components of the 
binary are identical. For such cases a binary system would appear 0.75 mag brighter 
than a single star. So, deviation of binary system cannot be higher than 0.75 mag on 
the brightness scale. In fact, deviation becomes smaller if one of the stars becomes 
cooler. When the cooler star's contribution becomes negligible, the system  
behaves as single star. Assuming that, we have various systems so that expected 
deviation would be about smaller than 0.40 mag on the brightness scale on an average. 
This limit is well within the standard deviation predicted in Table 3. Colour axis 
deviation from single stars are also expected to be negligible. Because at both 
limits colour of the system is like a single star. When two components were 
identical there would be no colour deviation. When the temperature difference 
between components increases system colour approaches to the colour of the dominant 
component.

The results of this study will help to understand space distributions and evolution 
of detached binaries in the solar neighbourhood. Indirectly, it will help to improve 
information about single stars. We suggest that the LCs relations are useful 
statistical tools to calculate the distances of detached binaries from their optical 
and near-infrared observations since the LCs relations are based on the most reliable 
distance estimation method (trigonometric parallax). Distances calculated from the 
LCs relations can give clues for astrometric observations of these systems, as well. 
Finally, it should be noted that future astrometric observations of detached 
binaries such as {\em GAIA} and {\em SIM} missions, will refine the LCs and PLCs 
relations.

\section{Acknowledgments}

We would like to thank the anonymous referee for his useful and constructive comments 
concerning the manuscript. This work has been supported in part by the Scientific 
and Technological Research Council (T\"UB\.ITAK) 106T688. This publication makes 
use of data products from the Two Micron All Sky Survey, which is a joint project 
of the University of Massachusetts and the Infrared Processing and Analysis 
Center/California Institute of Technology, funded by the National Aeronautics and 
Space Administration and the National Science Foundation. This research has made 
use of the SIMBAD database, operated at CDS, Strasbourg, France and NASA's 
Astrophysics Data System. Part of this work was supported by the Research Fund of 
the University of Istanbul, Project Number: BYP1379.

\end{document}